\documentclass[11pt]{article}
\pdfoutput=1
\usepackage{amsmath,amssymb,multirow,jheppub,graphicx}
\usepackage{centernot}
\usepackage{color}
\allowdisplaybreaks[1]
\usepackage{tikz}
 \tikzset{node distance=2cm, auto}
 
 \usepackage{hyperref}
\def\hhref#1{\href{http://arxiv.org/abs/#1}{#1}} 

\renewcommand{\Im}{\text{Im }}

\def\Im{\text{Im}}

\def\tr{\text{tr}}

\def\bar{\overline}

\def\I{{\cal I}}

\def\R{{\mathbb R}}
\def\coeff#1#2{{\textstyle {\frac {#1}{#2}}}}

\def\half{\coeff 12}
\def\N{{\cal N}}
\usepackage{verbatim}   
\usepackage[all]{xy}
\usepackage{bm}  
\def\Dslash{{\rlap{\raise 1pt \hbox{$\>/$}}D}}
\def\Pslash{{\rlap{\raise  1pt \hbox{$\>/$}}\,\partial}}

\title{
Deconstructing zero: resurgence, supersymmetry  and complex saddles 
      }
      \author[1]{Gerald V. Dunne}
\affiliation[1]{Department  of Physics, University  of Connecticut, Storrs, CT, 06269-3046}
\emailAdd{dunne@phys.uconn.edu} 
\author[2]{and Mithat  \"Unsal}
\affiliation[2]{Department  of Physics, North Carolina State University, Raleigh, NC, 27695}
\emailAdd{unsal.mithat@gmail.com}
\abstract{
We explain how a vanishing, or truncated, perturbative expansion, such as often arises in semi-classically tractable supersymmetric theories, can nevertheless be related to fluctuations about non-perturbative sectors via resurgence. We also demonstrate that, in the same class of theories,  
 the vanishing of the ground state energy (unbroken supersymmetry) can  be attributed to the cancellation between a real saddle and a complex saddle  
 (with hidden topological angle $\pi$),   and positivity  of the ground state energy  (broken supersymmetry) can  be interpreted as the dominance of complex saddles. 
 In either case,
 despite the fact that  the ground state energy is zero to all orders
in perturbation theory,   all orders of  fluctuations
around non-perturbative saddles are  encoded in the perturbative $E(N, g)$.
We illustrate these ideas with examples from supersymmetric quantum mechanics and quantum field theory.}

\begin{document}
\maketitle

\section{Introduction}
This paper addresses two puzzles concerning the interplay of supersymmetry (SUSY) and resurgence, aiming also for lessons of wider validity for non-supersymmetric theories.  We deconstruct two interesting ``zeros'' in SUSY theories,  revealing a hidden structure underneath.   

{\bf Puzzle one:}
It is well-known that in SUSY theories, either quantum field theory (QFT) or quantum mechanics (QM), the ground state energy is zero to all orders in perturbation theory 
\cite{Witten:1982df}:
\begin{align}
E_0^{\rm pert.}(g) = 0
\label{pert-1}
\end{align}
If SUSY is dynamically broken there are non-perturbative contributions of the form 
\begin{align}
 E_0^{\rm n.p.}(g)   = + g^\beta\,  e^{-A/g} \left( b_0 + b_1 g+ b_2 g^2 + \ldots \right)
 \label{NP-2}
 \end{align}
 At face value, this appears to be a counter-example to the resurgence idea  that perturbative data 
may be used to deduce non-perturbative saddles, and even the fluctuations around them \cite{Dunne:2016nmc}. In this case,  the perturbative  \eqref{pert-1} and non-perturbative data   \eqref{NP-2} appear to be unrelated. In Section \ref{sec:susy-qm} below, we show that resurgence does in fact connect the perturbative and non-perturbative fluctuations, in a subtle way.

{\bf Puzzle two:} 
 The  instanton  contribution  to the ground state energy is  {\it negative semi-definite} (in the absence of a topological theta angle or  a  Berry phase):
   \begin{align}
 \Delta E_0^{\rm instanton}(g)   \leq 0  
 \label{NP-inst}
 \end{align} 
On the other hand,  in SUSY theories the non-perturbative contribution to ground state energy is {\it positive semi-definite}:
      \begin{align}
 \Delta E_0^{\rm n.p.}(g) \geq 0
 \label{NP-susy}
 \end{align} 
 The opposite signs in (\ref{NP-inst}) and (\ref{NP-susy})  create a puzzle both in theories with or without  broken SUSY. However, resurgence, and its associated complex saddles, reconciles  \eqref{NP-inst} and \eqref{NP-susy}.

To begin, we recall two simple constraints coming from supersymmetry:
\begin{enumerate}
\item
In SUSY theories,  the SUSY algebra implies that the ground state energy is positive-semidefinite. Let $Q$ denote the supercharge,  $H$ the Hamiltonian, and  $|\Psi \rangle$ any physical state. Assuming  positivity of the norm of physical states,  
\begin{equation}
\langle H \rangle=  \half  \langle \Psi | \{Q^{\dagger}, Q\} | \Psi \rangle= \half  | Q  | \Psi \rangle|^2 + \half | Q^\dagger  | \Psi \rangle|^2  \geq 0
\end{equation}
Thus, the spectrum  is positive semi-definite. 
There is an obvious generalization to theories with  extended SUSY.  In particular, this means that the  ground state energy is  positive semi-definite. 

\item
A slight  refinement of this statement is the following: the ground state energy is zero to all orders in perturbation theory, 
and is   positive-semidefinite non-perturbatively: 
\begin{eqnarray}
E_0^{\rm pert.} (g) &=& 0
\label{susy-ineq1}\\
E_0^{\rm n.p.} (g)  &\geq& 0.
\label{susy-ineq}
\end{eqnarray} 
$E_0^{\rm n.p.}  >0$  amounts to dynamical breaking of SUSY, and 
$E_0^{\rm n.p.}  =0$  corresponds to unbroken SUSY \cite{Witten:1982df}. 
Here we explore how these relations fit with a semi-classical analysis.

\end{enumerate}

\section{Two types of resurgence in supersymmetric quantum mechanics  } 
\label{sec:susy-qm}

 In quantum mechanical systems, there are actually two types of resurgence. The first is generic, and the second is special to a class of systems:
 \begin{itemize}
\item[\it i)]
Conventional resurgence connects large orders of perturbation theory around a perturbative saddle to early terms of the perturbation theory around the first non-perturbative saddle in the topologically trivial sector \cite{Bogomolny:1980ur,ZinnJustin:1981dx}. Interestingly, this mimics closely the behavior of all-orders steepest descents expansions of ordinary contour integrals \cite{bh91}.

\item[\it ii)] 
It has recently been found that there is another type of resurgence, which is explicitly {\it constructive} \cite{Dunne:2013ada,Basar:2015xna,Dunne:2016qix,bdu}.  
For certain QM theories (including the paradigmatic `instanton' cases of the symmetric double-well, the periodic cosine potential, and the SUSY double-well), the fluctuations about each multi-instanton sector are encoded in the perturbative fluctuations about the vacuum sector. This encoding is  constructive in the sense that given some order of fluctuations about the perturbative vacuum, one can deduce the same order (minus one) of  fluctuations about any other non-perturbative sector.

\end{itemize}
In the following subsections we illustrate both types of resurgence for SUSY QM systems, with and without SUSY breaking.

\subsection{QM  with broken supersymmetry: the SUSY double-well}
\label{sec:susy-dw}

\subsubsection{{Resurgence (new constructive resurgence)} }
The simplest example of a QM system with dynamical SUSY breaking is the SUSY double-well, with superpotential  
\begin{align}
W(x) = \frac{x^3}{3} - \frac{x^2}{2} 
\end{align}
The Hamiltonian, in the non-perturbative normalization,  is  
\begin{align} 
H & =  \frac{g}{2} p^2 +  \frac{1}{2g} (W')^2 
   + \half (\psi^{\dagger}  \psi -1) W''  
\label{Ham-nf}
\end{align}
where $ \psi^{\dagger}  \psi $ is the fermion number operator, equivalently,  $\half (\psi^{\dagger}  \psi -1)= \half \sigma_3$ where $\sigma_3$ is the third Pauli matrix. 
As usual, 
projecting the Hamiltonian onto  fermion-number (spin) eigenstates,  leads to a pair of  purely bosonic graded Hamiltonians 
\begin{align} 
H_{\pm} & =  \frac{g}{2} p^2 +  \frac{1}{2g} (W')^2 
   \pm  \half  W''  
\label{Ham-graded}
\end{align}
Since the functions $e^{\pm \frac{W(x)}{g}}$, which are annihilated by the supercharge, 
are not normalizable, SUSY  is broken \cite{Witten:1982df}.  The ground state energy is zero to all orders in perturbation theory, while  the leading non-perturbatively induced ground state energy is \cite{Jentschura:2004jg}:
\begin{align}
E_0^{\rm pert.}(g) &= 0
\label{magic1-dw} \\
 E_0^{\rm n.p.} (g) &\sim  -\frac{1}{2\pi} e^{-\frac{S_b}{g}  + i\, \pi}  \left( 1   - \frac{5}{6}   g   - \frac{ 155  }{72}  g^2  
   - \frac{ 17315 }{1296}  g^3       - \frac{  3924815 }{31104} g^4 + O(g^5) \right) >0
   \label{magic2-dw}
\end{align}  
Here $S_b = 2 S_{ I} (= \frac{1}{3}$ with these normalizations) is the real part of the complex bion action, equal to twice the instanton action,  and $\pi$ is the imaginary part of the action of the complex bion, called the hidden topological angle \cite{Behtash:2015kna}.  We write the non-perturbatively induced correction, $ E_0^{\rm n.p.} (g) \sim   e^{-\frac{S_b}{g} }  >0$, as  $E_0^{\rm n.p.} (g) \sim -  e^{-\frac{S_{ b}}{g} + i \pi} $, to emphasize the fact that this contribution comes from a complex saddle, and it is
the existence of the associated hidden topological angle that resolves puzzle two, the apparent conflict between \eqref{NP-inst} and \eqref{NP-susy} \cite{Behtash:2015zha}.

To resolve puzzle one, note that perturbation theory for the energy spectrum in quantum mechanics is not only a function of the parameter $g$, but also of the unperturbed harmonic state's level number 
$N$: $E^{\rm pert}=E^{\rm pert} (N,g)$ \cite{Dunne:2013ada,Basar:2015xna,Dunne:2016qix}. Hidden in \eqref{magic1-dw} is the subscript zero, which labels the ground state level number  $N=0$. 
 The recent Mathematica Package ``BenderWu'' \cite{Sulejmanpasic:2016fwr}  by  Sulejmanpasic and one of us permits a simple computation, yielding in this model (see also \cite{Jentschura:2004jg}):
\begin{align}
E^{\rm pert.} (N, g)&= 
N - 3 N^2 g - \left(\frac{5}{2} N+ 17 N^3 \right) g^2 - \left(\frac{165}{2}  N^2 + \frac{375}{2} N^4\right)
    g^3  \cr 
&    - \left(\frac{1105}{8}  N  + \frac{9475}{4}  N^3   +\frac{10689}{4}   N^5 \right) g^4  +  O(g^5) 
\label{perturb}
 \end{align}
For the ground state, $N=0$, the perturbative energy  vanishes to all orders in $g$, as implied by SUSY.
 For other levels, the perturbative expansion (\ref{perturb}) is asymptotic,  with factorially growing non-alternating coefficients, together with a power $\frac{1}{ (2S_I)^k}$, at large orders $k$ in perturbation theory. 
 
 Since the perturbative coefficients in (\ref{perturb}) are polynomials in $N$, we can extend $N$ to a continuous parameter, with conventional Rayleigh-Schr\"odinger perturbation theory recovered at integer $N$. While $E^{\rm pert.} (N, g)$ vanishes at $N=0$, the same is not true for its derivative with respect to $N$:
 \begin{align}
\frac{\partial{E^{\rm pert.}}}{\partial N} &= 
 1 - 6 N g - \left(\frac{5}{2} + 51 N^2 \right) g^2 - \left(165 N + 750 N^3 \right) g^3  \cr 
 &- \left( \frac{1105}{8}    +  \frac{   28425}{4}  N^2  + \frac{53445}{4}   N^4 \right) g^4  + \ldots 
    \end{align}
   This series is asymptotic, with non-alternating factorially divergent coefficients, even at $N=0$.  
  
    We now show that  we can deduce the non-perturbatively induced ground state energy  in 
  \eqref{magic2-dw} from perturbation theory.  Write
  \begin{align}
 E_0^{\rm n.p.} (g) &\sim  \frac{1}{2 \pi} e^{-\frac{ S_b}{g} } { \cal P}_{\rm fluc}(N=0, g)\quad ,  \qquad S_b=2S_I= \frac{1}{3}
   \label{np-energy-DW}
\end{align}  
where ${ \cal P}_{\rm fluc}(N, g)$ is the fluctuation around the complex saddle.  The constant $S_b$ in expression (\ref{np-energy-DW})  is already present in the leading large-order asymptotics of the perturbative energy, and can be extracted easily. 
 The fluctuation series ${ \cal P}_{\rm fluc}(N, g)$ is also asymptotic.   Extending the results of \cite{Dunne:2013ada} to this model, one finds that 
 $ { \cal P}_{\rm fluc}(N,g)$ is  completely determined by $E^{\rm pert.} (N,g)$:
    \begin{align}
&{ \cal P}_{\rm fluc}(N, g) =  \frac{\partial{E^{\rm pert.}}}{\partial N}  \exp\left[ S_b  \int \frac{dg}{g^2} \left(  \frac{\partial{E^{\rm pert.}}}{\partial N}  - 1+ \frac{2Ng}{S_b} \right) \right]  
\label{magic-formula}
\end{align}
For the ground state,  $N=0$, Eqs (\ref{np-energy-DW}, \ref{magic-formula}) give the non-perturbative ground state energy purely in terms of perturbative data, in complete agreement with (\ref{magic2-dw}), which is obtained by an independent method in  \cite{Jentschura:2004jg}.
Thus, even though the ground state energy \eqref{magic1-dw} is zero to all orders in perturbation theory,  the perturbative expression, $E_{\rm pert} (N,g)$, encodes all the information around the non-perturbative complex saddle.  

We stress that this type of relation connecting early terms around the  perturbative vacuum saddle with early terms around a non-perturbative saddle  (in this case, a complex bion saddle) \cite{Dunne:2013ada,Basar:2015xna,Dunne:2016qix} is  quite distinct from the conventional resurgence, which connects the late terms of perturbation theory to early terms around a non-perturbative saddle \cite{Jentschura:2004jg}. 
     
\subsubsection{{Resurgence (traditional late term/early term resurgence)} } 

The perturbative ground state ($N=0$) energy vanishes (\ref{perturb}), which looks convergent, but in fact this "zero" is better understood as a cancellation between two identical formal divergent series, with a vanishing overall coefficient in the SUSY theory. We  probe this structure, and the associated `hidden' resurgence, by softly breaking SUSY, deforming the Yukawa-interaction in \eqref{Ham-nf} as
\begin{align}
\half W'' \rightarrow  \half \zeta W'' 
\label{deformation}
\end{align} 
With this perspective we recover both the standard ``late term/early term'' form of resurgence, and also the new constructive form of resurgence, even though these both look impossible from a naive inspection of the SUSY expressions (\ref{magic1-dw}, \ref{magic2-dw}).

The perturbative expansion coefficients now depend parametrically on $\zeta$ (see \cite{qes}):
\begin{eqnarray}
E^{\rm pert}(N, g; \zeta) &\sim& \sum_{n=0}^\infty a_n(N; \zeta) g^n 
\nonumber\\
&\sim& \left(N+\frac{1}{2}-\frac{\zeta}{2}\right)+\left(-\left[3N^2+3N+1\right]+\left[\frac{3}{2}+3N\right]\zeta-\frac{1}{2}\zeta^2\right)\, g \nonumber\\
&&\hskip -3cm +\left(-\left[17N^3+\frac{51}{2} N^2+\frac{35}{2}N+\frac{9}{2}\right]+\left[\frac{51}{2} N^2+\frac{51}{2} N+\frac{35}{4}\right] \zeta-\left[\frac{21}{2} N+\frac{21}{4}\right]\zeta^2+\zeta^3\right) g^2  \nonumber\\
&&\hskip -3cm +
\left(-\frac{1}{2}\left[375N^4+750N^3+792N^2+417N+89\right]+\frac{3}{2}\left[500N^3+750N^2+528N+139\right]\zeta \right. \nonumber\\
&&\hskip -3cm \left. -\frac{1}{2}\left[498N^2+498N+173\right]\zeta^2+\frac{123}{4}\left[2N+1\right]\zeta^3-4\zeta^4\right)g^3 +\dots
\label{eq:dw-zeta}
\end{eqnarray}
Perturbation theory is asymptotic for any  non-integer value of $\zeta$. We can use $\zeta$ to interpolate between SUSY and non-SUSY theories, and see how the resurgence properties evolve.  
 With this $\zeta$ deformation, the complex bion amplitude  and
its non-perturbative contribution to the ground state energy  are also modified to:
  \begin{align}
 E^{\rm n.p.}_{\pm}(N=0, g; \zeta)  \sim  [{\cal CB}]_{\pm} \sim  -\frac{1}{2\pi}  \left( \frac{g}{2} \right)^{\zeta -1} 
    \Gamma(\zeta )  e^{ \pm i \pi \zeta   } e^{-S_b/g}  { \cal P}_{\rm fluc}(N=0, g; \zeta)  
   \label{magic-5-dw}
\end{align}  
where ${ \cal P}_{\rm fluc}(N, g; \zeta)$ is the fluctuation  around the complex saddle. For the $N=0$ level, write:
\begin{align}
&{ \cal P}_{\rm fluc}(N=0, g; \zeta)  = b_0(\zeta) +   b_1( \zeta)    g  + b_2( \zeta)     g^2  +  b_3(\zeta)     g^3  + \cdots 
\label{c-b-2}
\end{align}
Here the $b_i(\zeta)$ are polynomials in $\zeta$, the first few of which are: 
\begin{align}\label{eq:TDW_bcoeff_DU}
 b_0 (\zeta) &=1 \\ 
 b_1(\zeta) &= \frac{1}{6}  \left(      -53 +   69  \zeta -  21 \zeta^2   \right)  \cr
  b_2(\zeta) &=   \frac{1}{72} \Big(-1277  - 78  \zeta + 2919   \zeta^2  -2160 \zeta^3 +   441 \zeta^4
\Big),  \cr
  b_3 (\zeta) &=  \frac{1}{1296} \left(-9261 \zeta ^6+44793 \zeta ^5-23166 \zeta ^4-133083 \zeta ^3+26370 \zeta
   ^2+413469 \zeta -336437\right)   \nonumber
\end{align} 
These polynomials are deduced from the perturbative expression (\ref{eq:dw-zeta}) using the relation (\ref{magic-formula}), which connects the perturbative and non-perturbative fluctuations.
Note that the result (\ref{magic-5-dw}) becomes multi-valued at any non-integer value of $\zeta$, due to the  $e^{ \pm i \pi \zeta   } $   factor,  but is single valued for integer $\zeta$. The physical reason for this is explained below.


A parametric  Bender-Wu analysis (using \cite{Sulejmanpasic:2016fwr}) permits extraction of the  large-order behavior of the $\zeta$-deformed perturbative coefficients $a_n(N; \zeta)$, including subleading corrections.  For the ground state energy of the $\zeta$-deformed theory one finds
\begin{align}
 a_{n}(N=0; \zeta)
 &\sim  -  \frac{1}{2\pi}   \frac{1}{(2 )^{\zeta -1} } \frac{1}{\Gamma(1 -\zeta)}  
   \frac{(n -\zeta )!} {(S_b)^{n-\zeta + 1}}  \cr&\qquad\times   \left[b_0(\zeta) +{ S_b b_1(\zeta)  \over (n-\zeta) }   +  {S_b^2b_2(\zeta)  \over (n-\zeta)(n-\zeta-1)}+ \ldots\right]   
 \label{BW-P-4}
\end{align}
Note the appearance of {\it the very same polynomials} $b_n(\zeta)$ that appear in the perturbative fluctuations about the complex saddle (\ref{c-b-2}, \ref{eq:TDW_bcoeff_DU}). This is remarkable: the $b_n(\zeta)$ are computed in two independent ways in the $\zeta$-deformed theory, from the perturbative/non-perturbative relation (\ref{magic-formula}) and also from the large-order growth of the perturbative coefficients, (This is done numerically, without any recourse with the dispersion relation),  and the result is completely consistent  with both forms of resurgence, even though the perturbative ground state energy vanishes at the SUSY point where $\zeta=1$. We make several comments:
\begin{enumerate}

\item
Not only do the large orders of the perturbative coefficients $a_n$ encode information about the low orders of the non-perturbative fluctuation coefficients $b_n$, but since the $b_n$ are determined explicitly by the $a_n$, this means that the high and low orders of the $a_n$ are themselves explicitly related.

\item 

For generic non-integer values of $\zeta$, the cancellation between the left/right Borel resummation of perturbation theory, $ {\cal S}_{\pm }  { E^{\rm pert}}  (N=0, g; \zeta)$, and the ambiguity in the imaginary part of the complex bion amplitude (\ref{magic-5-dw})  guarantees that the combination of perturbative and non-perturbative contributions is ambiguity free and meaningful: 
\begin{align} 
\Im \Big[   {\cal S}_{\pm }  { E^{\rm pert}}  (g)   + [{\cal CB}]_{\pm}   \Big] (g)\sim 0
\label{eq:generic}
\end{align}
This is the first of a tower of consistency conditions arising in a trans-series which is required to be real \cite{Aniceto:2013fka}. 

\item

At the SUSY point, $\zeta=1$, the perturbative expression for the ground state energy vanishes, and correspondingly the non-perturbative contribution  from the complex bion amplitude (\ref{magic-5-dw}) becomes ambiguity free, as the imaginary part vanishes: 
\begin{align}
 \Im   {\cal S}_{\pm }  {E^{\rm pert} }  & =0 \cr 
  \Im   [{\cal CB}]_{\pm}    & = 0  
\end{align}
Thus, (\ref{eq:generic}) is satisfied in a simple way in the SUSY theory.
The real part of the complex bion amplitude is unambiguous and gives the non-perturbative ground state energy \eqref{magic2-dw}, which has been computed by completely independent methods \cite{Jentschura:2004jg,Dunne:2013ada}. Note that the $b_i(\zeta)$ polynomials have a well defined $\zeta\to1$ limit. Despite the fact that perturbation theory becomes strictly zero and the non-perturbative complex bion amplitude becomes well-defined, the imprint of resurgence  is still present.  
A more refined way to view (\ref{magic1-dw}, \ref{magic2-dw}) is: 
\begin{eqnarray}
a_n  (N=0; \zeta=1) & \sim& 0   \times \frac{-1}{2\pi}  \frac{  (n-1)! } { (\frac{1}{3})^n}  \left( 1 -  \frac{  (\frac{1}{3}) \frac{5}{6} } {(n-1)} -   \frac{   (\frac{1}{3})^2    \frac{155}{72}  
 } {(n-1)(n-2)}  - \ldots \right)  
 \label{large-order-canc1}\\
\Im   [{\cal CB}]_{\pm}(\zeta=1) &\sim&  0 \times   \frac{1}{2}  e^{- (\frac{1}{3})/g} \;\;\;\;\; \Big(   \textstyle{ 1   - \frac{5}{6}   g   - \frac{ 155  }{72}  g^2  
   - O(g^3) } \Big)
   \label{large-order-canc2}
\end{eqnarray}
In this form, the traditional resurgence relation is manifest, noting the correspondence between the coefficients in (\ref{large-order-canc1}) and (\ref{large-order-canc2}), but multiplied by zero at the SUSY point.

\item

For $ \zeta= 0$, the system reduces to the symmetric double-well problem, and the ambiguous imaginary  part of the perturbative expansion cancels against the imaginary ambiguous part of 
the instanton/anti-instanton amplitude. The pole term in the complex bion amplitude (\ref{magic-5-dw})  is dropped to avoid double-counting, as this corresponds to uncorrelated 2-instanton events \cite{Behtash:2015zha}. A dispersion relation gives the correct large order behavior (including sub-leading terms) of perturbation theory in the bosonic theory, in  exact agreement  with  Eq.(101)  of the second paper in \cite{Dunne:2013ada},  computed by  a  different method. 

\item For higher integer values of $\zeta=M\in {\mathbb N}^+ \geq 2$, one finds that perturbation theory either truncates or is convergent, for the first $M$ energy levels \cite{qes}. For these levels, the generic factorial large order growth of the perturbative coefficients vanishes, due to an overall factor of $0$ coming from a gamma function factor in the higher $N$ version of (\ref{BW-P-4}). Correspondingly, the  non-perturbative contribution from the complex bion is unambiguous and real, for these levels. 
\end{enumerate}

\subsection{QM with unbroken supersymmetry: SUSY Sine-Gordon potential}
\label{sec:susy-sg}
 Consider a periodic superpotential, 
 \begin{align}
W(x) = \cos(x)  
\end{align}
Since $e^{\pm W(x)/g} $ are both normalizable on the circle $x \in[0, 2\pi]$, this  system has two SUSY ground states,  one bosonic and one fermionic, and so SUSY is unbroken \cite{Witten:1982df}.
The ground state energy is zero to all orders in perturbation theory, and  also vanishes
 non-perturbatively: 
 \begin{align}
E_0^{\rm pert.} (g) &= 0, \cr 
E_0^{\rm n.p.} (g)  &= 0.
\label{susy-eq}
\end{align} 
These two SUSY ``zeros'' actually hide a wealth of correspondences, which are revealed by a resurgence analysis.

\subsubsection{Resurgence (new constructive resurgence)} 
The perturbative energy  as a function of level number  and coupling can be computed using a Bender-Wu analysis  \cite{Sulejmanpasic:2016fwr,qes}:
\begin{align}
E^{\rm pert.} (N,g)&= 
N - \frac{N^2}{4}  g - \left(  \frac{N}{32}  +    \frac{N^3}{16}  
\right) g^2 - \left(\frac{7 N^2}{128} +\frac{5 N^4}{128}\right) 
    g^3  \cr 
&    - \left(   
\frac{25 N}{2048}+\frac{91 N^3}{1024}+ \frac{33 N^5}{1024}
 \right) g^4  +  O(g^5) 
\label{perturb-SG}
    \end{align}
This, of course, vanishes for the ground state, $N=0$, consistent with SUSY.
    
This perturbative data can be connected to non-perturbative data with the help of the perturbative-non-perturbative relation \eqref{magic-formula}. 
The difference from the SUSY double-well example, with broken SUSY, discussed in Section \ref{sec:susy-dw} is that in the SUSY Sine-Gordon system, with unbroken SUSY, there are two saddles contributing to the non-perturbative ground-state energy: a real bion and a complex bion \cite{Behtash:2015zha}.  The real bion reduces the ground state energy, while the complex bion increases it by exactly same amount, resulting in a cancellation:
\begin{align}
 E_0^{\rm n.p.} (g) &\sim 2[{\cal RB}] + 2 [{\cal CB}]_{\pm}    \sim   \frac{1}{\pi} \left( - e^{-\frac{S_b}{g}}  -   e^{-\frac{S_b}{g} \pm i \pi} 
 \right)    { \cal P}_{\rm fluc}(N=0, g)   =0 
 \label{eq:susy-sg-cancel}
 \end{align} 
 Note that this cancellation relies crucially on the complex action and quantized hidden topological angle  of the complex bion \cite{Behtash:2015zha}. Here $S_b= 2S_I (=4$ with these normalizations) is the real bion action,  also equal to the real part of the complex bion action. The fluctuations about each of the real and complex bion are equal, written in (\ref{eq:susy-sg-cancel}) as $ { \cal P}_{\rm fluc}(N=0, g)$. Despite the cancellation in (\ref{eq:susy-sg-cancel}), the fluctuations about {\it each} of the non-perturbative saddles can still be expressed in terms of the perturbative data using the perturbative/non-perturbative relation (\ref{magic-formula}):
\begin{align} 
 { \cal P}_{\rm fluc}(N=0, g) &\underbrace{=}_{ \eqref{magic-formula}}  \left(  1-\frac{g}{8}-\frac{3 g^2}{128}-\frac{13 g^3}{1024}-\frac{341
   g^4}{32768}+O\left(g^5\right) \right) 
   \label{magic-2}
\end{align}  
 
\subsubsection{{Resurgence (traditional late term/early term resurgence)} } 

As in the broken SUSY example of the SUSY double-well, discussed in Section \ref{sec:susy-dw}, for the unbroken SUSY example of the SUSY Sine-Gordon system, 
the perturbative ground state ($N=0$) energy vanishes, which at first sight looks convergent, but this ``zero'' is  better understood as a cancellation between two identical formal divergent series. As before, we probe this structure, and the associated `hidden' resurgence, by softly breaking SUSY, deforming the potential as in (\ref{deformation}).   
Then the perturbative expansion takes the form \cite{qes}
\begin{eqnarray}
E^{\rm pert}(N, g; \zeta) &\sim& \sum_{n=0}^\infty a_n(N; \zeta) g^n
\label{eq:an-sg} \\
&\sim & \left(N+\frac{1}{2}-\frac{\zeta}{2}\right) +\frac{1}{8}\left(-\left[2N^2+2N+1\right]
+\left[2N+1\right]\zeta\right) g \nonumber\\
&& +\frac{1}{64}\left(-\left[4N^3+6N^2+6N+2\right]+\left[6N^2+6N+3\right]\zeta- \left[2N+1\right] \zeta^2\right) g^2
\nonumber\\
&&\hskip -2cm +\frac{1}{256}\left(-\left[10N^4+20N^3+32N^2+22N+6\right]
+\left[20N^3+30N^2+32N+11\right]\zeta\right.
\nonumber\\
&&\hskip -2cm \left. -\left[12N^2+12N+6\right]\zeta^2 + \left[2N+1\right]\zeta^3\right)g^3+\dots
\label{eq:sg-zeta}
\end{eqnarray}
The non-perturbative  contributions to the ground state  energy come from both the real bion and the complex bion: 
\begin{align}
 E^{\rm n.p.}_{\pm} (N=0, g; \zeta) &\sim  2[{\cal RB}] + 2 [{\cal CB}]_{\pm}  \cr
& \sim   \frac{1}{\pi}   \left( \frac{g}{8} \right)^{\zeta-1}    \Gamma(\zeta ) 
 \left( - e^{-\frac{S_b}{g}}  -   e^{-\frac{S_b}{g} \pm i \zeta \pi} 
 \right) 
 { \cal P}_{\rm fluc}(N=0, g; \zeta)  
   \label{magic-5}
\end{align}  
Here ${ \cal P}_{\rm fluc}(N, g; \zeta)$ is the fluctuation around the real and complex saddle, which are again equal in the $\zeta$-deformed theory. These fluctuations can be computed from the perturbative expansion using the perturbative/non-perturbative relation (\ref{magic-formula}). For the ground state, write:
\begin{align}
&{ \cal P}_{\rm fluc}(N=0, g; \zeta)  = b_0(\zeta) +   b_1( \zeta)    g  + b_2( \zeta)     g^2  +  b_3(\zeta)     g^3  + \dots  \qquad
\label{fluc-3}
\end{align}
The first few $b_i(\zeta)$ polynomials are: 
\begin{align}\label{eq:DU_b_coeff_nu0}
 b_0(\zeta) &=1  \cr
 b_1(\zeta) &= \frac{1}{8}  \left(      -5 +   5  \zeta -  \zeta^2   \right)  \cr
  b_2(\zeta) &=   \frac{1}{128} \Big(-13  + 2  \zeta + 15  \zeta^2  -8  \zeta^3 +   \zeta^4
\Big),  \cr
  b_3(\zeta) &=   \frac{1}{3072} \Big( -\zeta ^6+9 \zeta ^5-10 \zeta ^4-51 \zeta ^3-10 \zeta ^2+381 \zeta -357 \Big)
\end{align}
As for the $\zeta$-deformed double-well potential of Section \ref{sec:susy-dw},  we can also use  a parametric Bender-Wu analysis  to find the large perturbative coefficients \cite{Sulejmanpasic:2016fwr,qes}. For the ground state:
\begin{align} 
  a_{n}(N=0; \zeta) & \sim - \frac{1}{\pi}   \frac{1}{(8)^{\zeta  -1} } \frac{1}{\Gamma(1 -\zeta)}  
   \frac{(n -\zeta )!} {(S_b)^{n-\zeta + 1}}\nonumber\\&\hspace{2cm}\times \left( b_0(\zeta) +  \frac{ (S_b)  \, b_1(\zeta) }{n-\zeta} +   \frac{ (S_b)^2  \, b_2(\zeta) }{(n-\zeta)(n-\zeta-1)} +    
   \dots  
   \right)    
\label{cat}
 \end{align}
Once again, notice the appearance of the very same polynomials $b_n(\zeta)$ that appear in the perturbative fluctuations about the complex saddle (\ref{fluc-3}, \ref{eq:DU_b_coeff_nu0}). 
We make several comments:
\begin{enumerate}

\item
Once again, because the $b_n$ are determined by the $a_n$, we see that the large orders and the low orders of the perturbative coefficients $a_n$ are themselves explicitly related.

\item
For generic non-integer $\zeta$, neither the perturbative nor the non-perturbative contribution to the ground state energy vanishes, and both forms of resurgence are clearly seen in the relations between the expressions (\ref{magic-5})--(\ref{cat}).

\item
For $\zeta=1$, the supersymmetric point,  the ambiguous imaginary part of the complex bion amplitude vanishes,  and perturbation theory for the ground state converges. 
Still, the information about the fluctuations around the complex bion is encoded in the vanishing perturbation theory. It is only the overall factor
$\frac{1}{\Gamma(1 -\zeta)} $ in (\ref{cat}) that vanishes, but the data in the 
$b_i (\zeta)$  polynomials is still there with a smooth $\zeta=1$ limit,  and  (\ref{fluc-3}, \ref{eq:DU_b_coeff_nu0}) reduce exactly to the fluctuation terms in \eqref{magic-2}.   This is the traditional resurgence at work, but slightly in disguise, being multiplied by an overall zero: 
\begin{align}
\label{large-order-canc}
a_n  & \sim 0   \times  \left(\frac{-1}{\pi}\right)   \frac{  (n-1)! } { (4)^n}  \left( 1 -  \frac{  (4) \frac{1}{8} } {n-1} -   \frac{   (4)^2    \frac{3}{128}  
 } {(n-1)(n-2)}  + \ldots \right)  \cr
\Im   (2[{\cal CB}]_{\pm} ) &\sim  0 \times    e^{- 4/g} \;\;\;\;\; \Big(   \textstyle{ 1   - \frac{1}{8}   g   - \frac{ 3  }{128}  g^2  
   - O(g^3) } \Big)
\end{align}
Note once again the correspondence between the expansion coefficients.
This is the realization of the traditional late-term/early term type resurgence at the supersymmetric point  $\zeta=1$. 

\item 

For $\zeta=0$, the large-order relation (\ref{cat}) reduces to that of the Mathieu system 
\cite{Dunne:2013ada,Basar:2015xna,Dunne:2016qix}, obtained by different techniques. 

\item
For integer values of $\zeta=M\in {\mathbb N}^+ \geq 2$, the perturbative expansion for the first $M$ levels either truncates
 or is convergent. This is consistent with the vanishing of the imaginary part of the non-perturbative saddle contribution in the higher $N$ version of (\ref{magic-5}), and corresponds to the existence of a nonlinear SUSY.

\end{enumerate}

\subsection{Summary of complex and real saddles in QM and the passage to QFT}
In the above SUSY QM examples, it turns out that the  {\it exact} complex and real  non-self-dual saddle solutions can be found explicitly \cite{Behtash:2015zha}.      The construction goes as follows:  after integrating out 
the fermions, one solves the complexified classical Euclidean equations of motion, in the inverted potentials $V_{\pm} = -\left(\frac{1}{2g} (W')^2   \pm  \half  W'' \right)$. A real saddle contribution to the ground state energy is negative,  $-e^{-S_b/g}$, while the contribution of a complex saddle is positive and of the form  $-e^{-S_b/g+ i \pi}$. These examples show  that the criterion for dynamical SUSY breaking reduces to a competition between real and complex saddles. If the complex saddles dominate, then SUSY is dynamically broken, and the ground state energy is positive. If the complex saddles and real saddles cancel each other, SUSY is unbroken and the ground state energy remains non-perturbatively  zero. Note that the real saddles never dominate, because that would lead to a negative ground state energy, which is impossible in a SUSY theory. In this sense, complex saddles in SUSY theories (and we expect also in non-SUSY theories) are more generic than real saddles. 

In QFT, one can formally integrate out the fermions, but it is much more difficult to construct exact non-BPS saddle solutions, real or complex. However, the QM examples in  \cite{Behtash:2015zha} also show that the exact saddles are in one-to-one corespondence with approximate bion solutions, which can be constructed in quantum field theory.  In the next Section we present several SUSY QFT examples in which approximate bion solutions play the role of exact saddle solutions, again resolving puzzles in the semi-classical analysis of SUSY theories.

\section{Complex Saddles in SUSY QFT}

In this Section we list some examples in SUSY QFT in which approximate real and complex bion solutions play the role of the real and complex saddles, and discuss their impact on a semi-classical analysis of SUSY.

\subsection{Theories with instantons and unbroken supersymmetry} 

Extracting the physical message from the QM examples above, we propose that in theories
without dynamical supersymmetry breaking, 
the vanishing of the ground state energy  be attributed to the cancellation between a real saddle  and a complex saddle with hidden topological angle $\Theta_h= \pi$.   In these cases, instantons never contribute to the ground state energy due to fermionic zero modes, and 
the leading contribution to vacuum energy starts at the second order in semi-classics, where it cancels due to an interference between real and complex saddle contributions:
 \begin{align}
 \label{np-vanish-2}
E_{\rm gr} \sim \left( \underbrace{ - e^{ -2S_I}}_{\rm real \; saddles}   \underbrace{- e^{ -2S_I \pm  i   \pi} }_{\rm complex\; saddles}  \right)\sim 0 
\end{align}

\subsection{Theories with instantons and broken supersymmetry} 
In theories with dynamical supersymmetry breaking, dynamical supersymmetry breaking can be re-phrased as the dominance of the complex saddles over the real saddles. 
In the simplest example of dynamical supersymmetry breaking, with only one complex saddle and no real saddle, the complex saddle contributes positively to the ground state energy (due to the hidden topological angle), and gives the non-perturbative contribution to the ground state energy: \begin{align}
 \label{np-positive}
E_{\rm gr} \sim  \left(  \underbrace{ -e^{ -2S_I \pm  i   \pi} }_{\rm complex\; saddles}  \right)> 0\, . 
\end{align}
The complex nature of the saddle is crucial for consistency of the semi-classical analysis with the positivity of the ground state energy.

\subsection{Runaway potentials in $\N=1$ SQCD in $d=4$}

Consider supersymmetric quantum chromodynmics (SCQD) with $N_f=N_c-1$, 
defined  on  $\R^4$. This system provides a special platform on $\mathbb R^4$ in which non-perturbative dynamics is calculable \cite{Affleck:1983mk}, as the usual problem with large instantons is resolved by the scalar meson vacuum expectation value 
 $\langle M  \rangle= \langle Q \tilde Q \rangle \sim v^2$.
Instantons with size larger than the inverse vev do not exist, and instanton calculus is reliable. 
 The instanton has $2N_c$ adjoint zero modes and $2N_f= 2N_c-2$ fundamental zero modes.  Due to Yukawa interactions of the form 
$Q\lambda \psi_Q$,  the instanton vertex can be converted to a diagram with only two fundamental fermion external lines,  $\psi_Q$ and  $\psi_{\tilde Q}$.  The instanton amplitude is proportional to $e^{-\frac{8 \pi^2}{g(\mu)^2} + i \theta}= \left( \frac{\Lambda}{\mu}\right)^{3N_c-N_f}=\left( \frac{\Lambda}{\mu}\right)^{2N_c +1} $, with an effective superpotential  \cite{Affleck:1983mk},
 \begin{align}
 \label{super-pot-3}
W (M)=  \frac{\Lambda^{2N_c+1}}{ \det M} 
\end{align}
The corresponding bosonic potential  can be interpreted as being due to 4d instanton/anti-instanton correlated pairs. This can be found either by using the superpotential, or alternatively, by integration over the quasi-zero mode "separation" between the the two instantons.

Since the  $[\I\bar \I]$ has negative quasi-zero modes,  the amplitude for the correlated event can be found by  integration over the separation Lefschetz thimble (or valleys in the terminology of \cite{Yung:1987zp}).  This 
 induces an extra phase $e^{i \pi}$, and the bosonic potential is 
 \begin{align}
 \label{super-pot-4}
V (v)=    -  \frac{e^{-2S_I + i \pi }  \mu^{4N_c+2}}  {v^{4N_c-2}  } \sim +   \frac{\Lambda^{4N_c+2}}{ v^{4N_c-2}} 
\end{align}
which is positive-definite.  In QM, the phase $e^{i \pi}$ arises as the imaginary part of the action of a complex saddle \cite{Behtash:2015zha}, which is the exact form of the $[\I\bar \I]$ event, living in the complexified field space. The positivity of the run-away potential can be understood from the semi-classical perspective as a consequence of the hidden topological angle of the complex bion. 
 
Adding a soft mass term for meson superfield changes the superpotential   \eqref{super-pot} into  \begin{align}
 \label{super-pot}
W (M)=  \tr \left(m M \right) + \frac{\Lambda^{2N_c+1}}{ \det M}  
\end{align}
This induces a mass for fermions, $m\psi_{Q_i} \psi_{\tilde Q_i}$, and the two fermionic zero modes of the instanton are lifted.  Consequently, the instantons also contribute to the bosonic potential. The associated bosonic potential is 
 \begin{align}
 \label{super-pot-1}
V (v_i) \propto    \sum_{i=1}^{N_c-1}  \left(   m^2  v_i^2 -  2 m  \frac{\Lambda^{2N_c+1}}{v_1^2 \ldots  v_{i}^2 \ldots    v_{N_c-1}^2}   +     \frac{\Lambda^{4N_c+2}}{ (v_1^2  v_2^2 \ldots   v_i^3 \ldots    v_{N_c-1}^2)^2    }  \right)
\end{align}
Setting $v_i=v, i=1, \ldots, N_c-1$,  and restoring the hidden topological angle associated with the $[\I\bar \I]$ bion event, it is instructive to rewrite this expression as 
 \begin{align}
 \label{super-pot-2}
V (v) \propto   \left(   m^2  v^2   \underbrace{ -2 m  \frac{\Lambda^{2N_c+1}}{v^{2N_c-2}} }_{\rm instanton \; (real  \; saddle)}   \underbrace{ -e^{i \pi}     
 \frac{\Lambda^{4N_c+2}}{ v^{4N_c-2} } }_{\rm complex \; saddle} \right)
\end{align}

There are a few interesting points to note about this bosonic potential. The first term is the tree level mass term, while the second and third terms are non-perturbatively induced. 
The two non-perturbatively induced terms have opposite signs. The instanton effect on the bosonic potential  is  negative, while the effect on the  potential due to the  $[\I\bar \I]$ event  is positive (due to the complex nature of the  $[\I\bar \I]$ saddle). Furthermore, the vanishing of the vacuum energy is due to a competition between the two positive terms (the tree-level term and the complex  $[\I\bar \I]$  saddle contribution) and the negative contribution from instantons. Minimizing the potential, we find a multi-branched result, $v^2=( \frac{\Lambda^{2N_c+1}}{m})^{1/N_c}$, related to the $N_c$-vacuum of the pure $\N=1$ SYM theory,  which can be obtained by integrating out the massive flavors.  Plugging this minimum value back into the potential, and 
writing $\Lambda^{2N_c+1}= e^{-S_I}$, we observe that the tree level term and 
complex saddle term give equal contributions, of the form $+ e^{-S_I/N_c}$, while the usual instanton term is given by  $- e^{-S_I/N_c}$. 
Thus the collective behavior of the combined complex saddles, instantons, and the tree level potential induce effects of fractional instanton type. Such configurations appear naturally on $\R^3 \times S^1$, however, they are mysterious  from the 4d point of view.

\subsection{Runaway potentials in  $\N=2$  Supersymmetric Yang-Mills (SYM)  in $d=3$}
 Similarly,  in three-dimensional ${\cal N}=2$ SUSY  gauge theory 
\cite{Affleck:1982as}, the bosonic potential can either be derived from 
the superpotential or by performing a quasi-zero mode integration, and 
provides a positive definite run-away potential,  
\begin{align}
V(\phi)  \sim  -e^{-2S_m \pm   i \pi}   e^{-2\phi}  \, . 
\end{align} 
The origin of the positive definiteness of the run-away potential is  
a complex phase that arises from the quasi-zero-mode integration for the associated complex saddle.

\subsection{Center symmetry in Yang-Mills, superYang-Mills and   QCD}

Consider  ${\cal N}=1$ SYM   theory compactified on $\R^3 \times S^1$. 
For small compactification radius, the gauge coupling is small  and a semi-classical approach is reliable, similar to the $N_f=N_c-1$ SQCD model on $\R^4$, discussed above. In both cases, the 4d instanton moduli are tamed, and instanton calculus is free of IR pathologies. 

 At  leading order in the semi-classical 
expansion,  this ${\cal N}=1$ SYM   theory on $\R^3 \times S^1$ has monopole-instantons which induce a superpotential (see for example 
\cite{Davies:2000nw}).   At second order in the semi-classical 
expansion, there are magnetic bions and neutral bions 
\cite{Unsal:2007jx,Anber:2011de,Poppitz:2011wy,Argyres:2012ka,Poppitz:2012sw}, in correspondence with the positive and negative entries of the extended Cartan matrix, respectively.  The bosonic potential for  the $SU(2)$ gauge theory 
is 
\begin{align}
V(\phi, \sigma)  \sim - e^{ -2S_m } \cos( 2 \sigma)  
  -e^{-2S_m \pm   i \pi}   \cosh(2\phi) \, ,
\end{align} 
where $\sigma$ is the dual photon field,  and $\phi$ is the  deviation of the holonomy field from the center-symmetric point. 
Note the complex phase in front of the neutral bion term, $e^{-2S_m \pm   i \pi} $, which guarantees that the potential  is bounded from below and has a minimum at $\phi=0$. The complex action arises from the integration over the quasi-zero mode thimble of the constituent monopole/anti-monopole pair, which is viewed as a complex saddle. 
Thus, the stability of the center symmetry in the SYM theory is due to the neutral bion, which is a complex saddle. An analogous situation occurs for 2D sigma models, such as $\mathbb C\mathbb P^{N-1}$ and Grassmanians, the principal chiral model and the $O(N)$ model \cite{Dunne:2012ae,Cherman:2013yfa,Dunne:2015ywa,Misumi:2016fno}.

Neutral bions also exist in Yang-Mills theory and QCD, and they are the leading non-perturbative effect to stabilize the center-symmetry, playing a vital role in the confinement-deconfinement phase transition \cite{Poppitz:2012sw,Liu:2015ufa}. This suggests that complex saddles should be important also in non-supersymmetric theories.

\section{Conclusions}

In this paper we have shown that resurgence may still relate the fluctuations about different perturbative and non-perturbative sectors, even when one or other of them vanishes, as often happens in a supersymmetric theory. We have illustrated these ideas with examples from SUSY quantum mechanics and SUSY quantum field theory. In the SUSY QM examples, we showed that both the generic `late term/early term' resurgence features, and the special constructive resurgence properties of the SUSY double-well and SUSY Sine-Gordon model are still present, even though certain ground state quantities vanish. In the SUSY QFT examples, the role of exact saddles is played by approximate  bion solutions.

It would be interesting to explore also localizable SUSY QFT, for which asymptotic and resurgence properties have recently been studied \cite{Russo:2012kj,Aniceto:2014hoa,Couso-Santamaria:2015wga,Honda:2016mvg,Gukov:2016njj,Marino:2016new}. As a simple analogy, consider the canonical example of localization in finite dimensional integrals, the exponential of the height function on $S^2$; this example is in one-to-one correspondence with SUSY QM. Generalizing to the height function in a particular direction on $S^d$, the corresponding integral
\begin{eqnarray}
\int d\Omega_{d-1} \int_0^\pi e^{i t \cos \theta} (\sin \theta)^{d-1}\, d\theta =\pi \left(\frac{2\pi}{t}\right)^{(d-1)/2}  J_{\frac{d-1}{2}}(t)
\label{eq:besselj}
\end{eqnarray}
can be evaluted as a Bessel function, with index depending on $d$.
Deforming $d$ away from the SUSY point $d=2$ is analogous to the soft SUSY breaking $\zeta$-deformation discussed in Section \ref{sec:susy-qm}. 
For arbitrary $d$, the integral localizes to the two critical points at $\theta=0$ and $\theta=\pi$, but in general the fluctuations about each $e^{\pm i t}$ term are asymptotic series, and these two asymptotic series are related by well-known resurgence relations.
When $d$ is an even integer these fluctuations truncate to polynomials, and therefore are convergent, analogous to the spectral properties of quasi-exactly soluble QM systems \cite{qes,Turbiner:1987nw}, which possess a nonlinear SUSY algebra \cite{Klishevich:2000dp}.
At the special (linear) SUSY point where $d=2$, each of the fluctuations has just one term. Nevertheless, for general $d$, resurgent asymptotics relates the fluctuations about the two saddles.

There are in fact many interesting physical examples where certain fluctuations vanish or truncate, but there are nevertheless associated non-perturbative contributions, related by a symmetry. Heat kernel and zeta function expansions generically involve asymptotic series, but simplify (and may even truncate) when there is a high degree of symmetry, such as for spheres \cite{minak}.
In matrix models and 2d gauge theory, certain weak coupling or large $N$ expansions may simplify or truncate, but there are still related physically interesting non-perturbative effects \cite{Gross:1980he,Wadia:2012fr,Witten:1992xu,Douglas:1993iia,Hatsuda:2013oxa,marcos-book}.
For the low energy effective action of Type IIB string theory in 10 dimensions, the perturbative expansion for graviton scattering has just two terms, but the imposition of U-duality, with its associated structure of automorphic forms, permits an explicit and unique non-perturbative completion \cite{Green:1997tv,Green:2010wi,Persson:2010ms}. We hope that the ideas in this paper may be useful to explore resurgent features of such systems.
\vskip 1cm

{\bf Acknowledgments}
We thank  C. Koz\c caz, D. Persson, B. Nielsen, T. Sulejmanpasic and  Y. Tanizaki for discussions.  
M.\"U's work was partially supported by the Center for Mathematical Sciences and Applications (CMSA) at Harvard University, where part of this work was done. 
This material is based upon work supported by the U.S. Department of Energy, Office of Science, Office of High Energy Physics under Award Number DE-SC0010339 (GD), and Office of Nuclear Physics under Award Number DE-SC0013036 (M\"U).

\end{document}